\documentclass{JINST}
\usepackage{graphicx, amsmath}

\title{Development of a Diehard GEM using PTFE insulator substrate}

\author{M. Wakabayashi$^a$\thanks{Corresponding author. Tel.: +81 48
    467 9336; fax: +81 48 462 4640.  E-mail address:
    masaki.wakabayashi@riken.jp (M. Wakabayashi).},
  K. Komiya$^b$, T. Tamagawa$^a$, Y. Takeuchi$^a$, K. Aoki$^a$, A. Taketani$^a$~and H. Hamagaki$^c$\\
  \llap{$^a$}RIKEN,\\
  Hirosawa 2-1, Wako-shi, Saitama 351-0198, Japan\\
  E-mail: \email{masaki.wakabayashi@riken.jp}\\
  \llap{$^b$}Tokyo Metropolitan Industrial Technology Research Institute (TIRI),\\
  Aomi 2-4-10, koto-ku, Tokyo 135-0064, Japan\\
  \llap{$^c$}Center for Nuclear Study (CNS), University of  Tokyo,\\
  Hirosawa 2-1, Wako-shi, Saitama 351-0198, Japan\\}

\abstract{We have developed the gas electron multiplier (GEM) using
  polytetrafluoroethylene (PTFE) insulator substrate
  (PTFE-GEM). Carbonization on insulator layer by discharges shorts
  the GEM electrodes, causing permanent breakdown. Since PTFE is hard
  to be carbonized against arc discharges, PTFE-GEM is expected to be
  robust against breakdown. Gains as high as 2.6$\times$10$^4$ were
  achieved with PTFE-GEM (50 $\mu$m thick) in Ar/CO$_{2}$ =
  70$\%$/30$\%$ gas mixture at V$_{\rm GEM}$= 730~V. PTFE-GEM never
  showed a permanent breakdown even after suffering more than 40000
  times discharges during the experiment. The result demonstrates that
  PTFE-GEM is really robust against discharges.  We conclude that PTFE
  is an excellent insulator material for the GEM productions.}


\begin{document}

\section{Introduction}

The gas electron multipliers (GEMs), which have been developed at CERN
by F. Sauli \cite{bib1}, are one of the micro-pattern gas detectors
that detect X-rays and charged particles with good spatial
resolution. We can easily manufacture a large size GEM detector with
lower cost than the same size of semiconductor detector, and we can
obtain better position resolution with GEM than with a scintillator
detector. Thus, GEMs have been used in many fields, such as particle
and nuclear physics, radiology, non-destructive testing and so on.

%GEMs have a structure that GEM foils are arranged in multistage
%between a cathode and a lead-out pad and it is detecting a charged
%particle by the gas amplification. 

For the ``standard GEM'', 50 $\mu$m thick copper coated Kapton
substrate is used. Typically, a number of through-holes with a
diameter of 70 $\mu$m are etched with a pitch of 140 $\mu$m into the
substrate \cite{bib1}. We also uses a CO$_2$ laser etching technique
for drilling the copper cladded polyimide (PI-GEM) \cite{bib4} or
liquid-crystal polymer (LCP-GEM) \cite{bib2}.

%The GEM foil shows an occasional breakdown by abnormal discharges
%under a large potential difference between electrodes to get higher
%gain. The resistive electrode GEM (RE-GEM) and Thick-GEMs were
%proposed and studied to reduce a chance of breakdown by the electric
%discharge \cite{bib5,bib6}. On the other hand, a statistical model
%predicting the GEM survival probability to discharge events is
%proposed by Cardini et al. \cite{bib7}. However, the perfect solution
%to prevent the GEM breakdown has not been proposed yet. In this paper,
%

High potential differences, which is necessary to achieve high gains,
can cause permanent damage by electric discharge. To handle this
problem, a statistical model predicting the GEM survival probability
to discharge events is proposed by Cardini et al. \cite{bib7}. On the
other hand, the resistive electrode GEM (RE-GEM) and Thick-GEMs were
proposed and studied to reduce a chance of breakdown by the electric
discharge \cite{bib5,bib6}. We consider a solution to this problem the
usage of new dielectrics as insulator for a GEM. Here we will present
our results of a new GEM using polytetrafluoroethylene (PTFE) as
insulator.

\section{The choice of an insulator material}

\subsection{Observation of a damaged standard GEM}

\begin{figure}[hbt]
  \begin{center}
    \includegraphics[width=12cm]{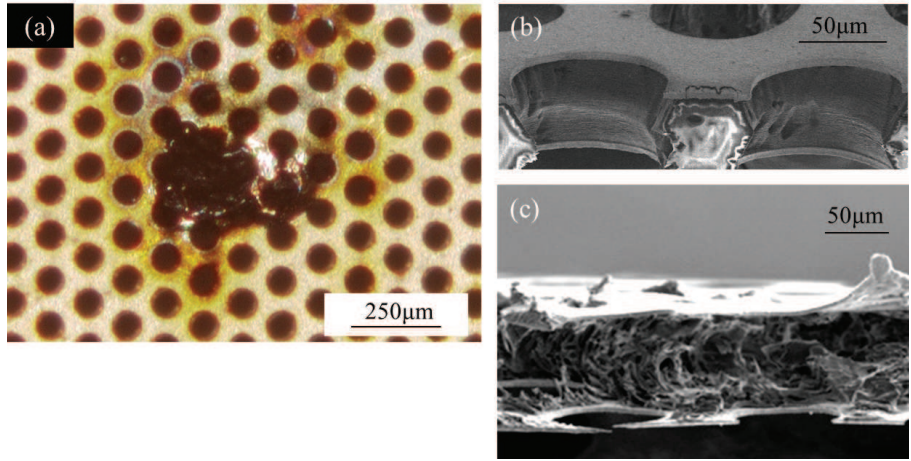}
    \caption{(a) A micrograph of the failure part of the PI-GEM.  (b)
      A SEM image of the normal part of the PI-GEM, and (c) that of
      the damaged part.}
    \label{fig01}
  \end{center}
\end{figure}

To investigate a primary risk to the GEM foil in operations, we have
analyzed a damaged GEM foil.  Figure \ref{fig01}(a) shows a top view
of the PI-GEM damaged by abnormal discharges.  The surface was
carbonized and the copper foil peeled off. Figure \ref{fig01}(b) shows
a scanning electron microscopic (SEM) image at the cross-section of
the normal part of the PI-GEM which was cut by an ultra-microtome. We
have demonstrated that the ultra-microtome technique was most suitable
method for obtaining the cross-section of the GEM foil
\cite{bib8}. Figure \ref{fig01}(c) shows a SEM image at the
cross-section of the damaged part. From those SEM images, we conclude
that the insulating layer changed into a porous body. Moreover, the
charging-up on the surface of the sample was not observed during the
SEM imaging, which normally occurs when the sample has a large surface
resistivity. Therefore we conclude that the resistivity around the
burned section was reduced due to carbonization of polymer.

Kapton, the standard insulator for GEMs, has the advantage of being a
commercial mass product and is widely used for the production of
flexible circuit boards. However, it has the disadvantage that it
easily carbonizes under the influence of the arc discharges during
operations. Thus, we thought that the robustness of the GEM may
improve if we can find a good insulating material which is not
carbonized by electrical discharges.

%An insulator material of GEM foils, which is generally PI or LCP, are
%commercial articles for flexible boards. The reason why those
%materials have been used is due to their availability and workability,
%although those materials are easy to be carbonized under the influence
%of 

\subsection{Arc resistance test}

To find a new material for the GEM insulator, we surveyed many
polymers and evaluated four of them by using an arc resistance test
method. The arc resistance is the ability of a material to resist the
influence of arc discharge. In the method, the arc resistance is
defined as the duration in seconds where the material resists the
formation of a surface-conducting path when subjected to an
intermittently occurring arc discharge of high voltage at a low
current characteristics \cite{bib9}. The thickness of a specimen in
the test was 3 mm by definition. Although the thermal conduction might
be quite different for thin foils, these tests is useful for
comparing different materials. Failure of the specimen may be caused
by heating to incandescence, burning, tracking or carbonization of the
surface.

%Breakdown between two tungsten electrodes usually occurs as a
%conducting path is burned on the surface of the dielectric material.
 
\begin{figure}[hbt]
  \begin{center}  
     \includegraphics[width=12cm]{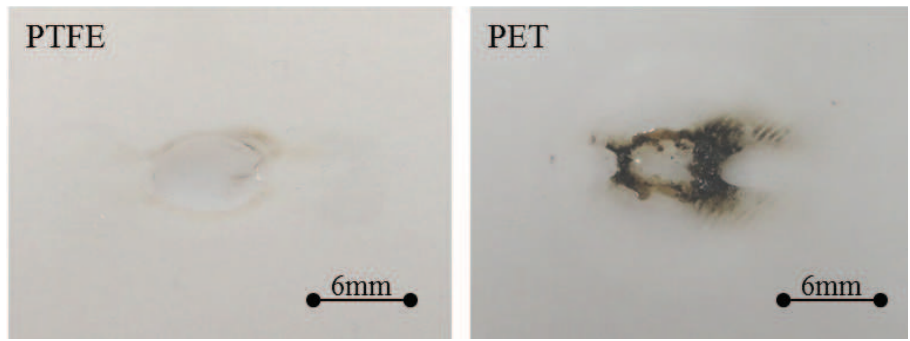}
     \caption{Photographs of the surface of the PTFE and PET specimens
       after the arc resistance tests.} 
     \label{fig02}
  \end{center}
\end{figure}

The characteristics of some well-known plastics, PI, LCP, polyethylene
terephthalate (PET), and PTFE are summarized in Table
\ref{tab1}. Among all the materials, PTFE has the highest value of the
arc resistance. We have evaluated the arc resistance with the test
equipment HAT-100 (a production of Hitachi Chemical Company, Ltd.) for
PTFE and PET. The test results are shown in figure \ref{fig02}. The
part of PET exposed by arc discharges was melted and burned, and a dip
like a crater was formed. The similar dip was seen on the surface of
PTFE, but there was no burnt deposit. To improve the lifetime of the
GEM foil against discharges, PTFE has the potential to be a suitable
material as insulating layer.

\begin{table}[hbt]
  \caption{Physical, mechanical, and electrical properties of some major plastics.}
   \label{tab1}
  \begin{center}
    \begin{tabular}{l c c c c}
      \hline
      materials & PI   & LCP  &PET     & PTFE  \\ \hline
      Density (g/cm$^{2}$) &1.43  &1.35  &1.4     &2.13-2.20 \\ \hline
      Tension strength (MPa) &315   &108   &48-73   &20-35 \\ \hline
      Water absorption ($\%$) &1.3   &0.08  &0.4     &0 \\ \hline
      Volume resistivity ($\Omega$cm) &10$^{17}$ &6$\times$10$^{16}$
      &10$^{17}$ &$>$10$^{18}$ \\ \hline
      Arc resistance (s)  &135 &186 &117 &$>$300 \\ \hline
      Melting point (deg C)               &-   &-   &258 &327 \\ \hline
     \end{tabular}
  \end{center}
\end{table}

\section{Production of PTFE-GEM}
\label{production_of_PTFE-GEM}
\subsection{Metal deposition on PTFE}

Since we could not find an appropriate commercial PTFE substrate which
was laminated with copper foils without any adhesive, we fabricated
the PTFE film with copper electrodes by ourselves. Copper was
sputtered on both sides of the 50 $\mu$m thick PTFE film, whose
thickness was selected because the standard CERN GEM uses the 50
$\mu$m thick PI, by the magnetron sputtering system SX-200 (ULVAC,
Inc.). The sputtering conditions were as follows: 1 kW of the applied
DC power, 0.67 Pa of the pressure, 120 mm of the length between the
target and the sample, 10 rpm for the rotation speed, and 22 min of
sputtering time. The sputtering was performed intermittently so that
the PTFE did not thermally transform by heat. The thickness of the
copper layer deposited on the PTFE film was 1.0 $\pm$ 0.2 $\mu$m
measured with a stylus profiler Dektak150 from Veeco Instruments Inc.

\subsection{Laser drilling through the PTFE substrate}

\begin{figure}[hbt]
  \begin{center}
    \includegraphics[width=12cm]{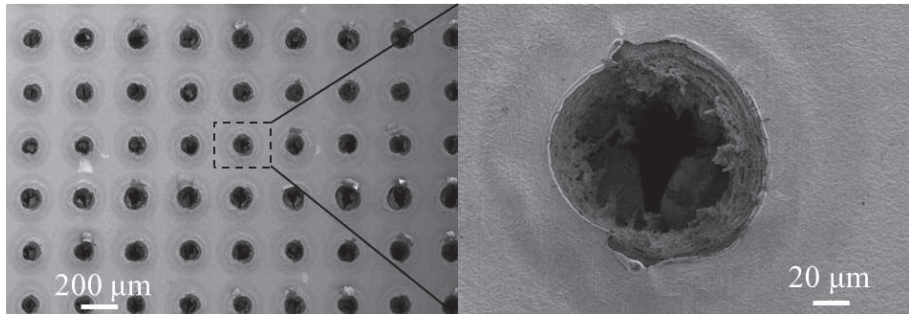}
    \caption{A SEM image of through holes drilled by CO$_2$ laser.}
    \label{fig03}
  \end{center}
\end{figure}

Methods such as chemical, plasma, and laser etching techniques are
used for drilling holes through the substrate of the GEM foil
production. Since PTFE has strong chemical resistance, the chemical
etching was hard to realize. First, we tried to drill the holes
through the copper-sputtered PTFE film using a CO$_2$ laser. Figure
\ref{fig03} shows a photo of the substrate drilled with a 30 W of
CO$_2$ laser. Since many debris remained inside the through-holes and
thermal damage by the laser could not be avoided, we gave up to use
the CO$_2$ laser for this purpose. In addition, the diameter of holes
became more than 100 $\mu$m (we expected around 70 $\mu$m.) and the
diameter was not controllable by our current method.

%The resolution of the detector improves so that the diameter of the
%holes are smaller and the density of the holes are higher.

Afterwards, we tried to drill the through-holes using a femtosecond
laser. By using this type of laser we can avoid heating-up the
substrate during the process. In addition, the multi-photon absorption
occurs under the extreme condition of the femtosecond laser, resulting
the efficient drilling as if an ultra-violet laser. The mean
wavelength of our femtosecond laser was 780~nm and the oscillation
frequency was 1~kHz. The laser beam was scanned by a galvano scanner,
and was focused on the surface of a sample with a telecentric lens of
100~mm focal length.

\begin{figure}[hbt]
  \begin{center}
     \includegraphics[width=12cm]{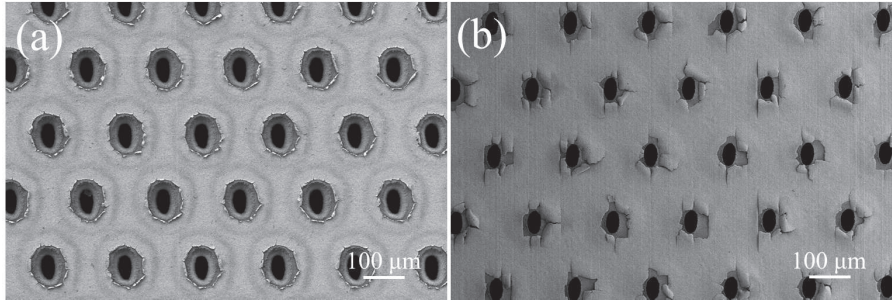}
     \caption{SEM images of PTFE-GEM drilled with the femtosecond
       laser: (a) Upper side (the surface irradiated by laser) and (b)
       lower side.}
     \label{fig04}
  \end{center}
\end{figure}

The copper-sputtered PTFE substrate we explained in the previous
section was used. The processing conditions were as follows: an output
power of 170~mW, a scanning speed of 0.5~mm/s, the pitch of zig-zag
arrangement of 200~$\mu$m, and a processing area of
20$\times$20~mm$^2$. When we drilled a hole, we moved the focal point
of the laser so that it created a circle with a diameter of 10~$\mu$m
around the center of the hole.  Figure \ref{fig04} shows the SEM
images of upper (irradiated side by the laser) and lower surfaces of
the PTFE substrate.

The shape of the through-holes was conical where the diameters of the
upper and lower sides were about 80 and 30 $\mu$m, respectively.
Since the beam spot of the femtosecond laser was not true-circle, the
shape of the holes became somewhat oval. In addition, the sputtered
copper was removed around the edge of the through-holes in both upper
and lower sides, and cracks on the copper were also seen. Those
defects are probably caused by the weakness of the film adhesion
between copper and PTFE, and the higher absorption rate of laser power
in copper than in PTFE.

The production quality of the PTFE-GEM foils inspected by eyes was not
good as we expected. However, we tested them if they work properly as
GEM because the resistance between the electrodes was more than
20~T$\Omega$, which was quite enough for operation. We left the
improvement on the production of the PTFE substrate and on the etching
methods for future works.

\section{Evaluation of PTFE-GEM}

\subsection{Experimental setup}

\begin{figure}[hbt]
  \begin{center}
    \includegraphics[width=12cm]{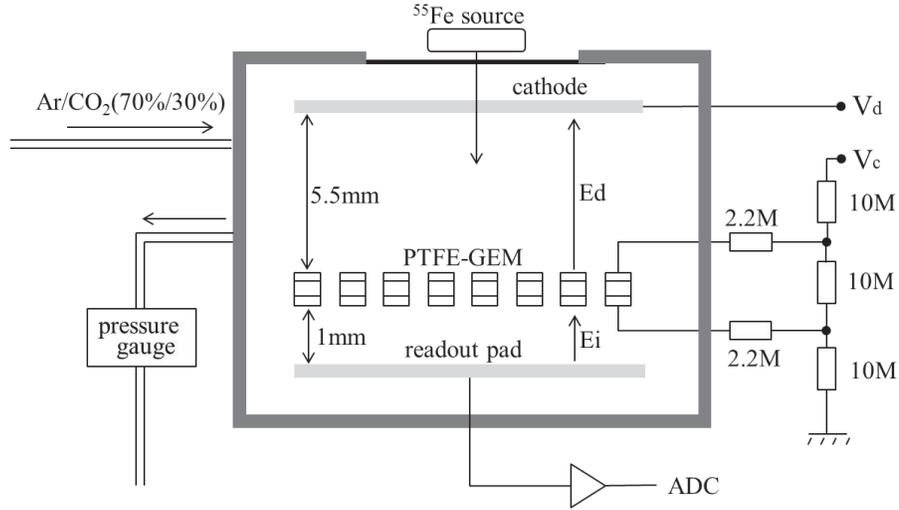}
    \caption{A schematic view of the GEM test setup.}
    \label{fig05}
  \end{center}
\end{figure}

Figure \ref{fig05} shows a schematic view of the GEM test setup used
in this study. The setup consisted of a cathode, a PTFE-GEM foil,
and a readout pad. PTFE-GEM was 50 $\mu$m thick with an active area of
20 $\times$ 20 mm$^{2}$ explained in 
\S\ref{production_of_PTFE-GEM}. The cathode was a 15 $\mu$m thick
aluminum foil. The cathode, PTFE-GEM, and readout pad were placed
in a chamber which was then filled with gas. The vertical space of the
target region, which was the space between the cathode and
PTFE-GEM, was 5.5 mm, and the induction region between PTFE-GEM and
the readout pad was 1.0 mm. A high voltage was applied via a chain of
10 M$\Omega$ resistors. To minimize the risk of electric surges, a 2.2
M$\Omega$ protection resistor was added in the series with each GEM
electrode. The electric field in the drift region was $E_{d}$ = 2.5 kV
cm$^{-1}$ and inside of the induction region was $E_{i}$ = 4-5 kV
cm$^{-1}$. During the test, Ar/CO$_2$=(70\%/30\%) gas was made to flow
through the system. Charge signals from the readout pads were fed into
an amplifier, then the signals were digitized by an VME-ADC module.

\subsection{Gain curves}

The effective gain ($G_{eff}$) of PTFE-GEM is given by 
\begin{equation}
G_{eff} = Const \times \cfrac{S_{mean}}{q_{e} n_{e}}
\end{equation}
$S_{mean}$ is the ADC peak value of incident monochromatic X-rays,
$q_e$ is the electron charge (1.602 $\times$ 10$^{-19}$ C), and $n_e$
is the number of electron-ion pairs created by an X-ray photon. A
typical value of n$_e$ is 212 for a 5.9 keV X-ray photon from the
radioactive $^{55}$Fe source in the Ar/CO$_{2}$=(70\%/30\%) gas
mixture \cite{bib10}. The constant value is derived from a calibration
curve obtained with a test pulse.

\begin{figure}[hbt]
 \begin{center}
   \includegraphics[width=12cm]{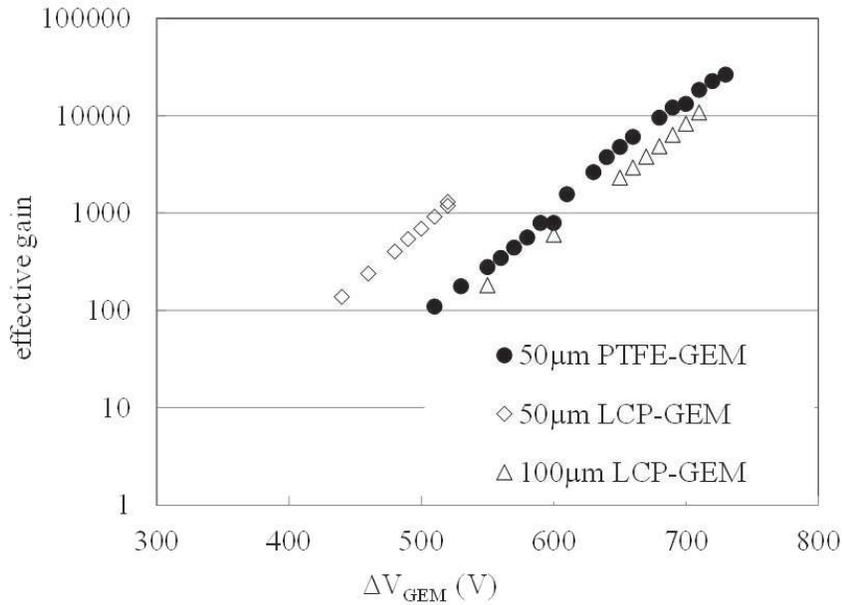}
   \caption{Effective gas gain of the 50 $\mu$m thick PTFE-GEM and the
     50 $\mu$m thick LCP-GEM and the 100 $\mu$ thick
     LCP-GEM.}
   \label{fig08}
 \end{center}
\end{figure}

Figure \ref{fig08} shows the gain curve of 50 $\mu$m thick PTFE-GEM.
The effective gain of PTFE-GEM we achieved was about 2.6$\times$10$^4$
at the applied voltage between PTFE-GEM electrodes of 730~V. We
stopped the gain curve measurement at this voltage because of the
limitation of the amplifier dynamic range. The gain curves for 50
$\mu$m thick LCP-GEM and 100 $\mu$m thick LCP-GEM are superposed in
the same figure for comparison. The gradient of 50 $\mu$m thick
PTFE-GEM was similar to that of 50 $\mu$m or 100 $\mu$m thick LCP-GEM.
The normalization of 50 $\mu$m thick PTFE-GEM was almost the same as
that of 100 $\mu$m thick LCP-GEM. Since the copper electrode receded
from the through-hole edge as shown in figure \ref{fig04}, the
amplification region affected by the path length and the electric
field strength might be different from the 50 $\mu$m thick LCP-GEM.

%effective length between the PTFE-GEM
%electrodes was about 100 $\mu$m, the field strength inside the
%through-holes might be weaken.

\subsection{Robustness against discharges}

\begin{figure}[hbt]
  \begin{center}
   \includegraphics[width=12cm]{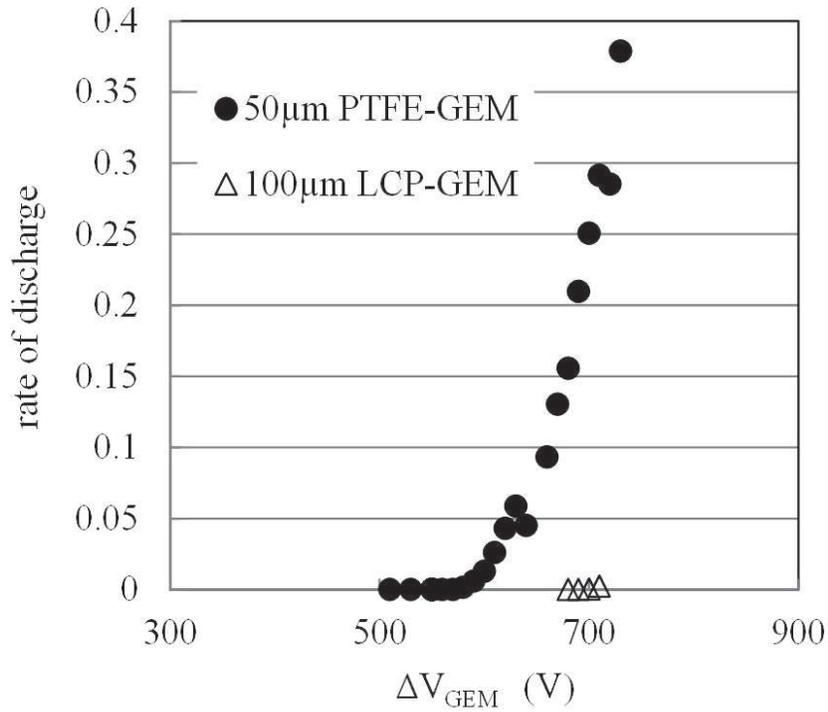}
   \caption{Discharge rate of the 50 $\mu$m thick PTFE-GEM and the 100
     $\mu$m thick LCP-GEM.}
   \label{fig06}
  \end{center}
\end{figure}

Figure \ref{fig06} shows the rate of discharges of PTFE-GEM as a
function of applied voltage between the GEM electrodes. For
comparison, the discharge rate of the 100 $\mu$m thick LCP-GEM is
superposed in the same figure.  The discharge rate of PTFE-GEM was
higher than that of LCP-GEM. This is probably due to the rough
structure around the edge of the through-holes (See figure
\ref{fig04}.). The roughness of the hole edge should be improved in
the next PTFE-GEM production.

The most important feature of PTFE-GEM against discharges is that it
never showed a permanent damage during the experiment. (The cumulative
number of discharges exceeded 4$\times$10$^4$ times.) On the other
hand, the LCP-GEM suffered permanent breakdown after 91 discharges at
an applied voltage of 710 V. We were convinced that the PTFE-GEM is
really robust against discharges. We would like to stress that we are
not suggesting to operate PTFE-GEM under continuous discharges, this
should be avoided for this kind of GEM as also for the standard GEM,
but we demonstrated that the PTFE is very robust against discharges.

%Here we note that we did not intend
%to operate PTFE-GEM under discharges. Not only for the standard GEM
%but also for PTFE-GEM, discharges should be avoided. In this
%subsection, we just demonstrated that PTFE-GEM was robust enough
%against discharges.

\begin{figure}[hbt]
  \begin{center} 
    \includegraphics[width=12cm]{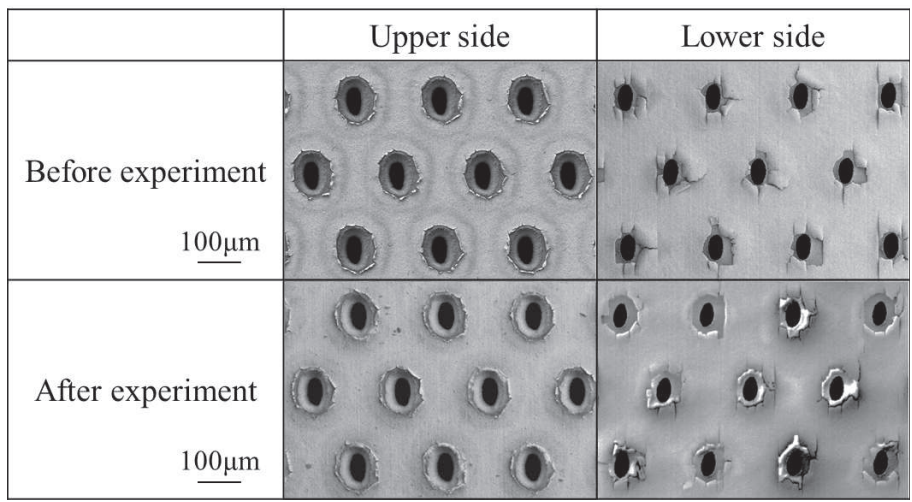}
    \caption{SEM images PTFE-GEM for before and after the
      experiment. Note that the SEM images do not shows the same
      region.}
    \label{fig09}
  \end{center}
\end{figure}

After the experiment, we have inspected the effective area of PTFE-GEM
with SEM. In figure \ref{fig09} the SEM images of PTFE-GEM for before
and after the experiment are shown. Across the entire region of
PTFE-GEM no carbonization, melting, or porous body was observed. As
conclusion of our experiment, PTFE-GEM has superior characteristic
that was not affected by discharges and PTFE is one of the best
insulator materials for GEMs due to this fact although further work is
necessary to optimize the production process.

\section{Summary and Outlook}

We have searched the insulator material which is hard to carbonize
when it is exposed by discharges. Judging from the arc resistant
tests, we selected PTFE as an insulator material for our GEM foil
production. We have produced the PTFE-GEM foils for the first time: we
formed copper electrodes on both sides of the PTFE film by magnetron
sputtering, and then drilled through-holes by using a femtosecond
laser processing technique.

We have measured the effective gain of PTFE-GEM. The gain we achieved
was about 2.6$\times$10$^4$ for the 50 $\mu$m thick PTFE-GEM in
Ar/CO$_2$=(70\%/30\%) gas mixture at an applied voltage of 730~V
between the PTFE-GEM electrodes. The gain curve was similar to that of
the 100~$\mu$m thick LCP-GEM. The discharge rate of PTFE-GEM was quite
high probably because of the roughness of the edges around the
through-holes. However, PTFE-GEM was never destroyed despite of the
fact that it suffered more than 40 thousand discharges during the
experiment. We have experimentally confirmed that PTFE-GEM is really
robust against discharges. No defect was observed on the PTFE-GEM
surface with SEM after the experiment.

As conclusion of our experiment, PTFE-GEM has superior characteristic
that was not affected by discharges. PTFE is one of the best insulator
materials for the GEM production although further work is necessary to
optimize the production process. To improve the production process, we
are fabricating the PTFE substrate with thicker foil electrodes, since
the thicker electrode can minimize the damage by heat in drilling the
holes. Another possible improvement is to use an UV laser for drilling
the PTFE substrate. A high energy laser may be easily punch the
substrate without heat damage.

We have an outlook to use PTFE-GEM for space applications such as
photoelectron tracking X-ray polarimeters \cite{bib11, bib12}. The
polarimeters are basically a coupling of single layer GEM to a fine
pixel readout chip. Using a stack of GEMs introduces an additional
smearing which limits the imaging capability of the system, while one
needs high gains at the same time. To achieve a good imaging
performance, the robustness of PTFE-GEM against discharges is crucial.

\acknowledgments

We thank Dr. K. Sunouchi and Dr. I. Takahashi at RIKEN for their help
to use the femtosecond laser. We thank Dr. K. Fujiwara and
Mr. T. Kobayashi at Tokyo Metropolitan Industrial Technology Research
Institute (TIRI) for their discussions and comments. This work was
partially supported by JSPS KAKENHI Grant Number 22244034.


\begin{thebibliography}{9}

\bibitem{bib1}
F. Sauli, \emph{Nucl. Instr. and Meth.}, A386, (1997) 531.

\bibitem{bib4}
T. Tamagawa, et al., \emph{Nucl. Instr. and Meth.}, A560, (2006) 418.

\bibitem{bib2}
T. Tamagawa, et al., \emph{Nucl. Instr. and Meth.}, A608, (2009) 390.

%\bibitem{bib3}
%M. Inuzuka, et al, \emph{Nucl. Instr. and Meth.},A 525, (2004) 529.

\bibitem{bib7}
A. Cardini, et al., \emph{IEEE Nuclear Science Symposium Conference Record}, (2005) 1127-1131. 

\bibitem{bib5}
A. Yoshikawa, et al., \emph{Journal of Instrumentation.}, Vol.7, Issue06, (2012) C06006.

\bibitem{bib6}
A.Breskin, et al., \emph{Nucl. Instr. and Meth.}, A598, (2009) 107.

\bibitem{bib8}
K. Komiya, et al., \emph{RIKEN Accelerator Progress Report}, 45, (2012) 170. 

\bibitem{bib9}
ASTM Stand., (2004) D495-99. 

\bibitem{bib10}
F. Sauri, \emph{CERN Yellow Report}, (1997) 77-09. 

\bibitem{bib11}
R. Bellazzini, et al., \emph{Nucl. Instr. and Meth.}, A560, (2006) 425.

\bibitem{bib12}
J.K. Black, et al., \emph{Nucl. Instr. and Meth.}, A581, (2007) 755.

\end{thebibliography}
\end{document}